\documentclass{ws-p8-50x6-00}

\newcommand\D{\ensuremath{{\cal D\,}}{}}
\newcommand{\DWI}{\ensuremath{{\cal D}_\textrm{\scriptsize Wi}\,}{}}
\newcommand{\DFP}{\ensuremath{{\cal D}_\textrm{\scriptsize Fp}\,}{}}
\newcommand{\DOV}{\ensuremath{{\cal D}_\textrm{\scriptsize Ne}\,}{}}
\newcommand{\unitmatrix}{\textrm{\boldmath{$\mathsf{1}$}}}

\begin{document}

\title{On the Spectrum of Lattice Dirac Operators}

\author{C. B. Lang}

\address{Institut f{\"u}r Theoretische Physik, 
Karl-Franzens-Universit{\"a}t Graz,\\
A-8010 Graz, AUSTRIA,\\E-mail: christian.lang@kfunigraz.ac.at}

\maketitle

\abstracts{With the Schwinger model as example I discuss properties
of lattice Dirac operators, with some emphasis on Monte Carlo
results for topological charge, chiral fermions and eigenvalue spectra.}

%%%%%%%%%%%%%%%%%%%%%%%%%%%%%%%%%%%%%%%%%%%%%%%%%%%%%%%%%%%%%%
\section{Introduction}\footnotetext{Talk at the
Int. Conf. on Mathematical Physics and Stochastic
Analysis (In honor of Ludwig Streit's 60th birthday), Lisbon, Oct. 1998}

%%%%%%%%%%%%%%%%%%%%%%%%%%%%%%%%%%%%%%%%%%%%%%%%%%%%%%%%%%%%%%
%%%%%%%%%%%%%%%%%%%%%%%%%%%%%%%%%%%%%%%%%%%%%%%%%%%%%%%%%%%%%%
\subsection{Continuum Concepts vs. Lattice Concepts}
%%%%%%%%%%%%%%%%%%%%%%%%%%%%%%%%%%%%%%%%%%%%%%%%%%%%%%%%%%%%%%

Relativistic particle physics is described by relativistic quantum field
theory; the theories explaining the fundamental interactions are gauge
theories.  QFT as it is formulated in terms of Lagrangians and functional
integration has to be regularized. The only known regularization scheme that
retains the gauge symmetry is replacement of the space-time continuum by a
space-time lattice. For convenience and other reasons this is done in an
Euclidean world. 

The most prominent QFT is QCD, the SU(3) gauge theory of quarks and  gluons.
There various non-perturbative phenomena appear intertwined: confinement and
chiral symmetry breaking. The classical continuum gauge fields $A$ that are
continuous and differentiable, living on compact manifolds, may be classified
by a topological quantum number (the Pontryagin index) $Q(A)$. Changing through
continuous deformations of the field  from one such topologically defined
sector to another is impossible. 

Topology is closely related to the fermion zero modes; these are eigenstates
(eigenvalue $E = 0$) of the Dirac operator
\be
i\gamma_\mu \left( \partial_\mu - i\,e\, A_\mu\right) \psi
 =  E \psi = 0 \quad  \rightarrow \quad\gamma_5\,\psi=\pm \, \psi\;.
\ee
Due to the anti-commutation of $\gamma_5$ with $\D$ the zero modes can be
chosen as eigenstates  of $\gamma_5$ with definite chirality. The Atiyah-Singer
index theorem (ASIT)\cite{AtSi71} relates the topological charge $Q(A)$ of the
background gauge field to these modes,
\be
Q(A)=\textrm{index}(A)\equiv n_{+} -n_{-}\;.
\ee
where $n_\pm$ denotes the number of independent zero modes with positive or
negative chirality, respectively. In two dimensions, another theorem of the
continuum -- the so-called Vanishing Theorem\cite{Ki77NiSc77An77} -- ensures
that only either  positive or negative chirality zero modes occur. 

Quantization involves summation over non-differentiable fields and the
lattice formulation does not provide for a unique definition of topological
charge at all. Any lattice definition involves implicit or explicit
assumptions, usually on the continuity and smoothness at the scale below one
lattice spacing. Quantization usually means Monte Carlo integration on finite
lattices and it is legitimate to question the ergodicity of the respective
``simulation'' with regard to the topological sectors.

Even the task of putting fermions on the lattice introduces limitations.
Various lattice Dirac operators have been proposed. It has been demonstrated
early\cite{NiNi} that within quite general assumptions (like locality and
reflection positivity) it is not possible to have single chiral fermions;
chiral symmetry has to be broken explicitly. For the simple Wilson action, the
breaking of the symmetry is so bad that no trace of the chiral properties of
the continuum theory is kept in the lattice theory. If, however, chiral
symmetry is broken already on the level of the action, how can one hope to
identify the (expected) spontaneous chiral symmetry breaking of the full
theory? Is there a lattice version of the ASIT? As we will discuss below,
recent developments do allow us to construct lattice Dirac operators, which
break chiral symmetry in a minimal way. 

Confinement, on the other hand, proved to be more straightforward. The gauge
coupling $g$ enters the lattice gauge action in form of the multiplicative
coupling $\beta=1/g^2$. The lattice formulation works particularly well at
strong coupling (small $\beta$), where confinement within a non-vanishing
domain of the couplings was proved. In the Monte Carlo calculations for lattice
QCD up to now no signal was found that indicates a phase transition between
that confinement phase and the  weak coupling (perturbative) regime; for finite
temperature such a deconfinement transition was established. 

In order to retrieve continuum QFT numbers for the physical quantities -- like
masses of hadronic bound states or certain matrix elements -- one has to show
that all dimensional physical quantities scale according to the scaling
function of the lattice spacing $a(\beta)$, which in turn should asymptotically
(in the continuum limit $\beta\to\infty$, $a\to 0$) agree with continuum
renormalization group scaling. 

Current lattice studies have to live in that environment: Try to  improve
scaling properties in order to get reliable continuum  results and try to deal
with chirality without destroying it from the begin!

%%%%%%%%%%%%%%%%%%%%%%%%%%%%%%%%%%%%%%%%%%%%%%%%%%%%%%%%%%%%%%
\subsection{Lattice Dirac Operators}\label{sec:GenProperties}
%%%%%%%%%%%%%%%%%%%%%%%%%%%%%%%%%%%%%%%%%%%%%%%%%%%%%%%%%%%%%%

All Dirac operators have $\gamma_5$-hermiticity,
\be
\gamma_5 \,\D\,\gamma_5=\D^\dagger\;.
\ee
It follows that
\begin{itemize}
\item The eigenvalues are either real or are complex conjugate pairs.
\item The operator $\gamma_5\,\D$ has a real spectrum.
\item We denote by $v_i$ the eigenvector of $\D$ for eigenvalue $\lambda_i$;
then $\gamma_5\,v_i$ is an eigenvector of $\D^\dagger$ for the same
eigenvalue. 
\item The diagonal entries of the chiral density matrix vanish for
non-real eigenvalues: 
$\lambda_i\notin\textrm{Re}\rightarrow\langle v_i|\gamma_5
\,v_i\rangle=0$.
\item The non-diagonal entries of the chiral density matrix vanish
whenever their respective eigenvalues are not complex conjugate pairs:
$\lambda_i\ne\bar\lambda_j\rightarrow\langle v_i|\gamma_5
\,v_j\rangle=0$.
\end{itemize}
An important conclusion is that only real eigenvalues lead to
contributions to the diagonal elements. Thus these modes 
are the only candidates for zero modes in the continuum limit:
in the continuum theory the (normalized) zero modes contribute to the diagonal
entries, $\langle \psi|\gamma_5\,\psi\rangle=\pm 1$.

Chirality is necessarily broken for lattice Dirac operators. However, barely
noticed for almost two decades, Ginsparg and Wilson\cite{GiWi82} formulated a
condition (GWC) under which circumstances\cite{Ha98c} remnants of the chiral
symmetry  survive in a lattice action for massless fermions. If the Dirac
operator obeys
\be
\frac{1}{2}\,\left\{\gamma_5,\D\right\} = a \, \D\,\gamma_5\,R\,\D\;,
\ee
where $R$ denotes a local matrix, then chiral symmetry is violated only by a
local term ${\cal O}(a)$. Dirac operators satisfying the GWC will be called GW
Dirac operators.

L\"uscher\cite{Lu98} has pointed out the explicit form of the associated
symmetry of the action,
\be
\psi \; \rightarrow \;
\exp\left[i\,\theta\,\gamma_5\,(1-a\,R\,\D)\right]\,\psi\;,\;\;
\bar\psi \; \rightarrow \;
\bar\psi\,\exp\left[i\,\theta\,(1-a\,\D\,R)\,\gamma_5\right]\;.
\ee

The GWC provides a sensible way (respecting the Nielsen-Ninomiya
theorem) to construct suitable Dirac operators. As has been
shown\cite{Ho98} such actions cannot be ultra-local, but
may be local, i.e. the coupling exponentially damped in real space.

For certain classes of Dirac operators one may derive stringent bounds on the
shape of the eigenvalue spectrum. This has been discussed in the framework of
so-called fixed point (FP) actions\cite{HaNi94,Ha98c}.

On the lattice real space renormalization group (RG) transformations
are realized through block spin transformations. The field variables
over a localized region around a site $x$ are averaged
to produce the field variable of a coarser lattice. This mapping procedure
in the space of configuration ensembles may be associated to a 
parameter flow in the space of actions. It is implicitly assumed
that indeed a Gibbsian measure is suitable to describe the ensemble of 
blocked configurations.

\begin{figure}[t]
\begin{center}
\epsfxsize=5cm 
\epsfbox{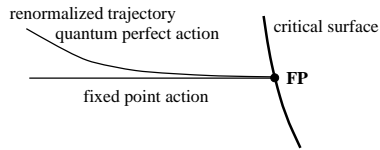} 
\end{center}
\caption{The FP action is the classical perfect action; the 
quantum perfect action may deviate from it away from the FP. \label{fig:FPflow}}
\end{figure}

The continuum limit of the quantum theory is obtained at a critical FP of the
system (cf. Fig.\ref{fig:FPflow}).  The so-called renormalized trajectory in
this space of actions leads to the FP and along its path there are no
corrections to scaling. Ideally one could simulate the system  at any point of
the renormalized trajectory (with the so-called quantum perfect action) and
obtain ``perfect''  continuum results. In reality that action may be quite
complicated and has to be truncated in the number of coupling constants. 

Hasenfratz et al.\cite{HaNi94} have suggested to determine instead the action
at the FP of the RG transformation. For (asymptotically) free theories it
may be determined from the  classical field equations. The action has
therefore been baptized  ``classical perfect action''.  Its classical 
predictions agree with those of the continuum action independent of the
coarseness of the lattice. FP actions are solutions of the GWC \cite{Ha98c}.

\begin{figure}[t]
\begin{center}
\epsfxsize=4cm
\epsfbox{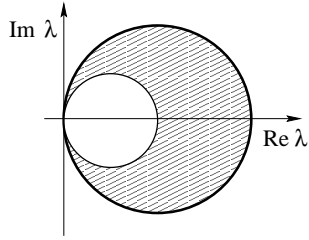} 
\end{center}%
\caption{For FP Dirac operators the eigenvalues are confined in the
shaded area of two circles tangential to the imaginary axis.
The thick line indicates the position of the spectrum for $R=1/2$.
\label{fig:FPSpectrumBounds}}
\end{figure}

At the classical level, for FP actions, the Atiyah-Singer theorem finds
correspondence on the lattice\cite{Ha98c,HaLaNi98};  at the quantum
level, no fine tuning, mixing and current renormalization  occur, and a
natural definition for an order parameter of the  spontaneous breaking
of the chiral symmetry is possible\cite{Ha98a}.
$R$ is then local and bounded and as a consequence 
the spectrum of ${\cal D}$ in complex space is confined between two
circles\cite{HaLaNi98} (cf. Fig.\ref{fig:FPSpectrumBounds}),
\be\label{CirclesBound}
|\lambda-r_{min}|\geq r_{min} \;,\quad |\lambda-r_{max}|\leq r_{max}\;\;,
\ee
where the real numbers $r_{min}$ and $r_{max}$ are related to the maximum
and minimum  eigenvalue of $R$ respectively. For non-overlapping BSTs
$R=1/2$ and (\ref{CirclesBound}) reduces to $|\lambda-1|=1$, i.e.
the spectrum lies on a unit circle.

Independent implementations of the GWC are provided by the overlap 
formalism\cite{NaNe}, which allows the formulation  of
chiral fermions on the lattice. These solutions are obtained in an elegant
way, as shown recently by Neuberger\cite{Ne98},  through some map
of the Wilson operator with negative  fermion mass. In this case we have
$R=1/2$ and circular spectrum $|\lambda-1|=1$, too.

For $R=1/2$ the GWC assumes the simple form
\be
\D\,+\,\D^{\dagger} = \D^{\dagger}\,\D = \D\,\D^{\dagger}\;.
\label{eq:gwc}
\ee

On the lattice, an index for $\D$ may be defined in a way analogous  to the
continuum, explicitly expressed by the relation\cite{HaLaNi98}
\be\label{index}
\textrm{index}(U)=-\textrm{tr}(\gamma^5 \,R\, \D)\;.
\ee
For GW Dirac operators and $R=1/2$ this relation comes out 
trivially. Only the
modes with real non-vanishing eigenvalues contribute to the trace in
the r.h.s; since the overall chirality must be zero, it
reproduces up to a sign $\textrm{index}(U)$. 
This index can be used to define a fermionic lattice
topological charge, $Q_{\rm ferm}(U)\equiv\textrm{index}(U)$, 
for which the ASIT is satisfied by definition. {\em For the FP action} that
fermionic definition coincides\cite{HaLaNi98}  with  the pure-gauge
quantity $Q_\textrm{\scriptsize Fp}(U)$, the FP topological 
charge\cite{BlBuHa96} of the
configuration $U$:
\be\label{index_fp}
Q_{\rm ferm}(U)\equiv\textrm{index}(U) = Q_\textrm{\scriptsize Fp}(U)\;.
\ee
The non-obviousness of this relation relies on the fact that  $Q_{\rm
Fp}(U)$ can be defined in the pure gauge theory,  without any regard to the
fermion part. This result is particular for a FP action and has no counterpart
for a general (non-FP) GW action. Of course, in practical implementations
one relies on approximate parametrizations of the FP Dirac operator and the
strictness of the relation is lost.

%%%%%%%%%%%%%%%%%%%%%%%%%%%%%%%%%%%%%%%%%%%%%%%%%%%%%%%%%%%%%%
\subsection{Test Bed Schwinger Model}
%%%%%%%%%%%%%%%%%%%%%%%%%%%%%%%%%%%%%%%%%%%%%%%%%%%%%%%%%%%%%%

As our test bed we consider a 2-dimensional (2D) 
quantum field theory with U(1) gauge 
group and $N_f$ flavors of fermions. The action for the massless
continuum model reads 
\be
S = \int d^2 x\, \left(\frac{1}{4}\,F_{\mu\nu}\,F_{\mu\nu}
+\sum_{f}^{N_f}\bar\psi_f\,\D\,\psi_f\right)\;.
\ee
For $N_f=1$ this
is the Schwinger model\cite{Sc}. 

This 2D version of QED resembles 4-dimensional QCD in various 
ways\cite{SM,GaSe94}. 
Quarks are ``trapped'', i.e. in a mechanism superficially mimicking
confinement we observe only bosonic asymptotic states. For $N_f=1$ this is the
Schwinger boson (called $\eta$ by analogy to 4D) with the mass
$m_\eta=g/\pi$ for the physical gauge coupling $g$. For the 2-flavor
model one expects also a triplet of massless bosons (pions)\cite{GaSe94},
although in 2D there can be not spontaneous symmetry breaking due to the
Mermin-Wagner-Coleman theorem\cite{MeWaCo}. Finally, there is a
non-vanishing condensate $\langle\bar\psi\,\psi\rangle$ due to an anomaly.

In the lattice formulation the gauge action in the compact
Wilson-formulation is written 
\be
S_g=\beta\,\sum_{p\in\Lambda} \,\left(1-\textrm{Re}\, U_p\right)\;,
\ee
where $\Lambda$ denotes the lattice $Z_N^2$ and the plaquette variable $U_p$ 
is
the oriented
product of links variables $U_{x,\mu}\in U(1)$ at site $x$ in
direction $\mu=1,2$. In the continuum limit $U_{x,\mu}\simeq
\exp{(i\,a\,g\,A_{x,\mu})}$ where $A$ is the gauge field in the
non-compact continuum formulation, $a$ denotes the lattice spacing, and
$\beta=1/(g^2\,a^2)$.

A lattice version for the integer geometric topological charge of the gauge 
field may be defined,
\be
Q(A) = \frac{g}{2\pi}\int d^2x\,  F_{12}(x)\quad\rightarrow\quad
Q(U) = \frac{1}{2 \pi}\sum_p \textrm{Im} \,\textrm{ln}\, U_p \;.
\ee
In torus geometry\cite{JoSaWi} (i.e. periodic boundary conditions) 
this number may be non-zero for compact gauge field configurations.

The lattice fermion action is formally
\be\bar\psi\,\D(m)\,\psi\;,
\ee 
where $\D(m)$ denotes the lattice Dirac operator matrix  (fermion mass $m$)
and the fermions are Grassmann fields. In a 2D context the Dirac matrices
$\gamma_\mu$ and $\gamma_5$ are to be replaced by $\sigma_\mu$ and $\sigma_3$.

In the quantized theory the Grassmann integration over the fermions yields
factors  $(\det \D)$ and the expectation value of some operator $C$
may be written
\be
\langle C \rangle = \frac{1}{Z}\,\int[dU]\,e^{S_g(U)}\,
\left(\det \D\right)^{N_f}\,C(U)\;,
\ee
where one usually samples over the gauge fields with some Monte Carlo
procedure. In the so-called {\em quenched} approximation ($N_f=0$) the
fermionic determinant is not included; this essentially neglects the fermionic
vacuum loops. 

{\em Dynamical} fermions can be included either by incorporating
them into the sampling probability measure for the gauge configurations
or in the observable. The first case is realized in the so-called
``Hybrid Monte Carlo'' method. However, since the determinant may be
negative, only even numbers of mass-degenerate fermions can be studied
then. The second approach requires calculation of the determinant for
each gauge configuration. It is plagued by the notorious sign problem,
giving rise to possibly violent fluctuations and large statistical
errors of the results.

The gauge integral should sample over all topological sectors. For
Dirac operators which allow exact zero modes the determinant weight
(for massless fermions) removes the corresponding contribution to the integral.

%%%%%%%%%%%%%%%%%%%%%%%%%%%%%%%%%%%%%%%%%%%%%%%%%%%%%%%%%%%%%%
\section{Lattice Fermions and Topology}
%%%%%%%%%%%%%%%%%%%%%%%%%%%%%%%%%%%%%%%%%%%%%%%%%%%%%%%%%%%%%%

We have been studying the lattice Schwinger model for various lattice sizes 
and values of the gauge coupling, both quenched and with dynamical fermions.
Here we discuss some of our results, with some  emphasis on the properties of
the spectra of GW operators.

%%%%%%%%%%%%%%%%%%%%%%%%%%%%%%%%%%%%%%%%%%%%%%%%%%%%%%%%%%%%%%
\subsection{The Wilson Dirac Operator}
%%%%%%%%%%%%%%%%%%%%%%%%%%%%%%%%%%%%%%%%%%%%%%%%%%%%%%%%%%%%%%

The original Wilson action\cite{Wi74} for fermions has the form
\be\label{DWI}
\DWI=(m+2)\,\unitmatrix-\frac{1}{2}\,M
\ee
with the hopping matrix
\be \label{hopM}
M_{xy}= \sum_{\mu=1,2} \left[
 (1+\sigma_\mu) \,U_{xy}\,\delta_{x,y-\mu}
+ (1-\sigma_\mu)\, U_{yx}^\dagger\,\delta_{x,y+\mu}\right]\;.
\ee
The hopping parameter $\kappa$ defines the fermion mass, for 
free lattice fermions $\kappa=1/(2\,m +4)$. 
However, for the 
full theory with gauge interactions this relationship no longer 
holds. On one hand due to quark trapping there are no asymptotic fermions.
On the other hand chiral symmetry is explicitly broken.

Since this breaking is a local feature, we expect that at some 
$\kappa_{c}(\beta)$
the chiral symmetry is formally restored in the sense, that the corresponding
Ward identity is satisfied with a vanishing fermion mass parameter.
This idea\cite{BoMaMa85} has been pursued in QCD\cite{JaLiLu96}
and we have utilized it also in the Schwinger model with Wilson fermions
in order to identify the position of the critical line, where
the fermion mass parameter vanishes\cite{HiLaTe98}.

\begin{figure}[t]
\begin{center}
\epsfxsize=8cm 
\epsfbox{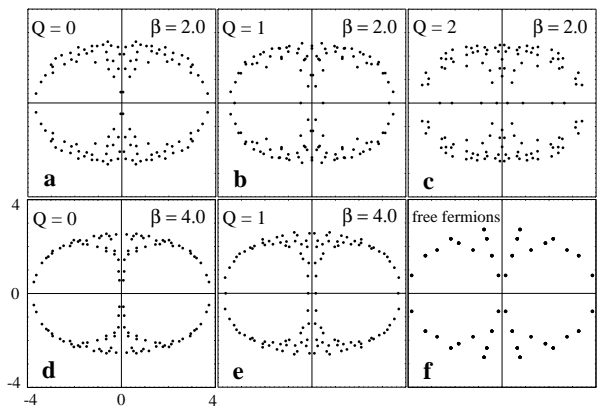} 
\end{center}
\caption{Spectrum of the hopping matrix $M$ for an $8\times 8$ lattice at various 
values of $\beta$, close to $\kappa_c(\beta)$; $Q$ denotes
the (geometric definition of the) topological charge. The eigenvalues for
free fermions are degenerate.\label{fig:WilsonSpectra}}
\end{figure}

In Fig.\ref{fig:WilsonSpectra} we compare\cite{GaHiLa97a} typical eigenvalue
spectra of the hopping matrix $M$ at different values of the gauge coupling and
in different topological sectors. We find strong correlation between the number
of real eigenvalues and the topological charge. Closer inspection leads to the
conclusion, that indeed the real modes  (counted according to their 
chirality\cite{He98}) may be interpreted as the would-be zero modes. Actually
one has to divide that number by a trivial multiplicity factor of 4, since only
the rightmost  modes become  zero modes of $\DWI$, whereas the other ones are
doublers at other corners of the Brillouin zone.  The agreement of this number
with the topological charge improves towards the continuum limit. 

In that sense even the Wilson Dirac operator obeys the
ASIT for $\beta\to\infty$; e.g. at $\beta=4$ in 99\% of the
gauge configurations (sampled in a hybrid Monte Carlo simulation with
dynamical fermions) the numbers of real modes (divided by 4) agrees with the
topological  charge. Also the Vanishing theorem is obeyed\cite{GaHiLa97a}.

%%%%%%%%%%%%%%%%%%%%%%%%%%%%%%%%%%%%%%%%%%%%%%%%%%%%%%%%%%%%%%
\subsection{The Fixed Point Dirac Operator}\label{FPsect}
%%%%%%%%%%%%%%%%%%%%%%%%%%%%%%%%%%%%%%%%%%%%%%%%%%%%%%%%%%%%%%

Although it was possible to obtain FP actions for some scalar or fermionic
systems (cf. the summary in Ref.5) it turned out to be be a formidable problem
for realistic gauge-fermion systems like QCD. We have  succeeded\cite{LaPa98b}
to find explicitly an approximate FP action for the  lattice Schwinger model.
The FP Dirac operator is parametrized by a set of 429 terms bilinear  in the
fermion and anti-fermion fields, 
\be
\DFP=\sum_{i,x,s} \,\rho_i(f)\,\sigma_i\,U(x,f)\,\delta_{y,x+f}\;.
\ee
The sum runs over paths $f$ connecting a central site $x$ with sites in a 
$7\times 7$ neighborhood; $U(x,f)$ denotes the product of link gauge variables
along that path. Standard symmetries are taken into account. Altogether we
considered 123 independent coupling parameters.

The general idea is to start with a parametrization of the action, generate a
set of gauge field configurations, perform the real space  RG transformation
and identify the blocked action. The  matrix elements of the Dirac operator on
the blocked configurations are then compared with the parametrization and the
coupling constants adjusted. The whole procedure is iterated until the
parameters converge.

We studied samples of 50 gauge configurations and $14\times14$ lattices at
large values of $\beta$.  We could show, that the resulting fermion action had
couplings damped  exponentially with their spatial extent; although the
parametrized action is anyhow ultra-local by construction, this observation
indicates that locality of an untruncated FP action seems feasible.

Using the FP action we determined the bosonic bound state propagators for the
$N_f=1$ and 2 Schwinger model and found excellent rotational invariance and
good scaling properties (cf. Fig.\ref{fig:PiEtaProp}), to be discussed below in
context of other actions.

\begin{figure}[tb]
\begin{center}
\epsfxsize=5cm 
\epsfbox{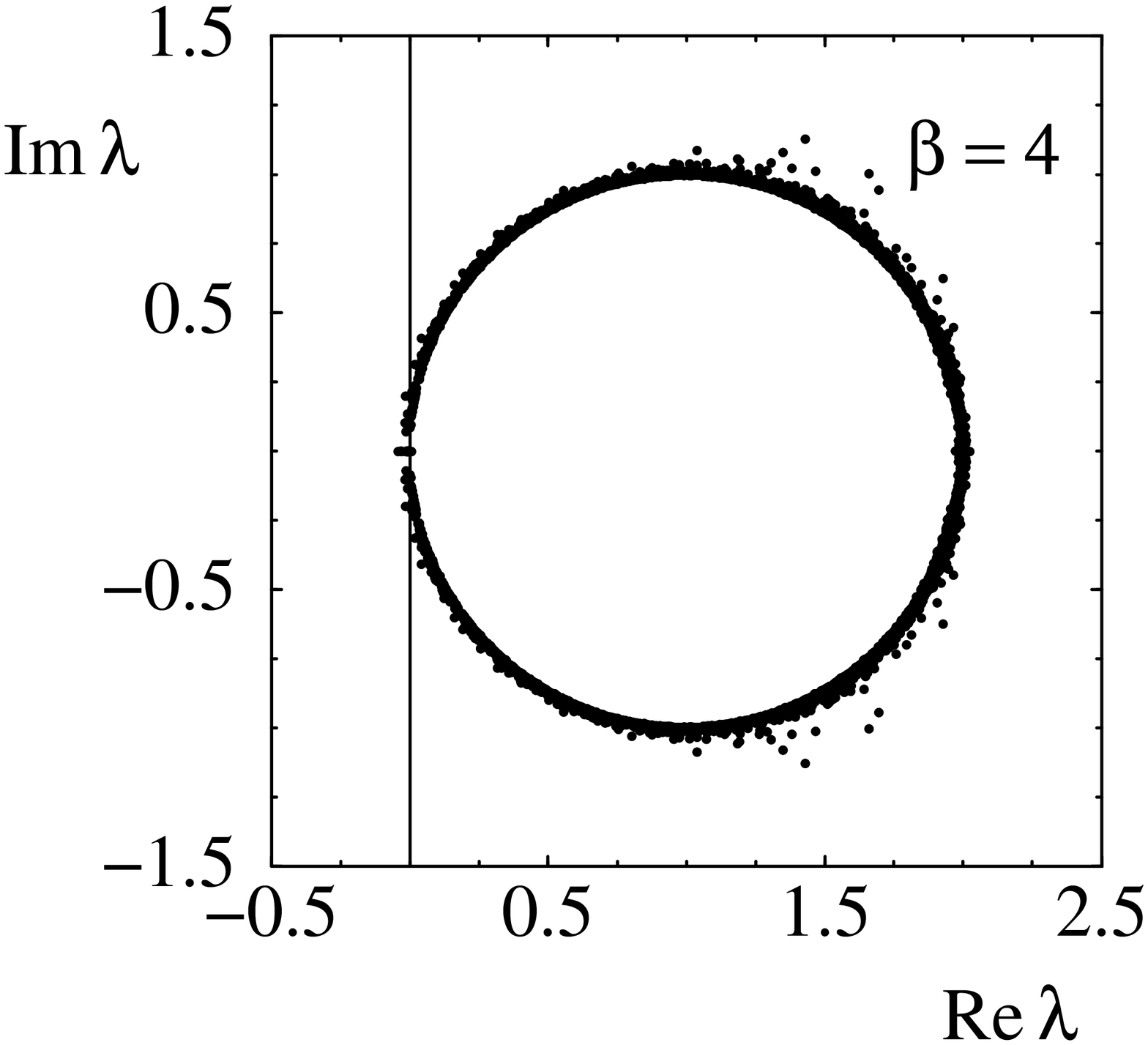}
\epsfxsize=5cm
\epsfbox{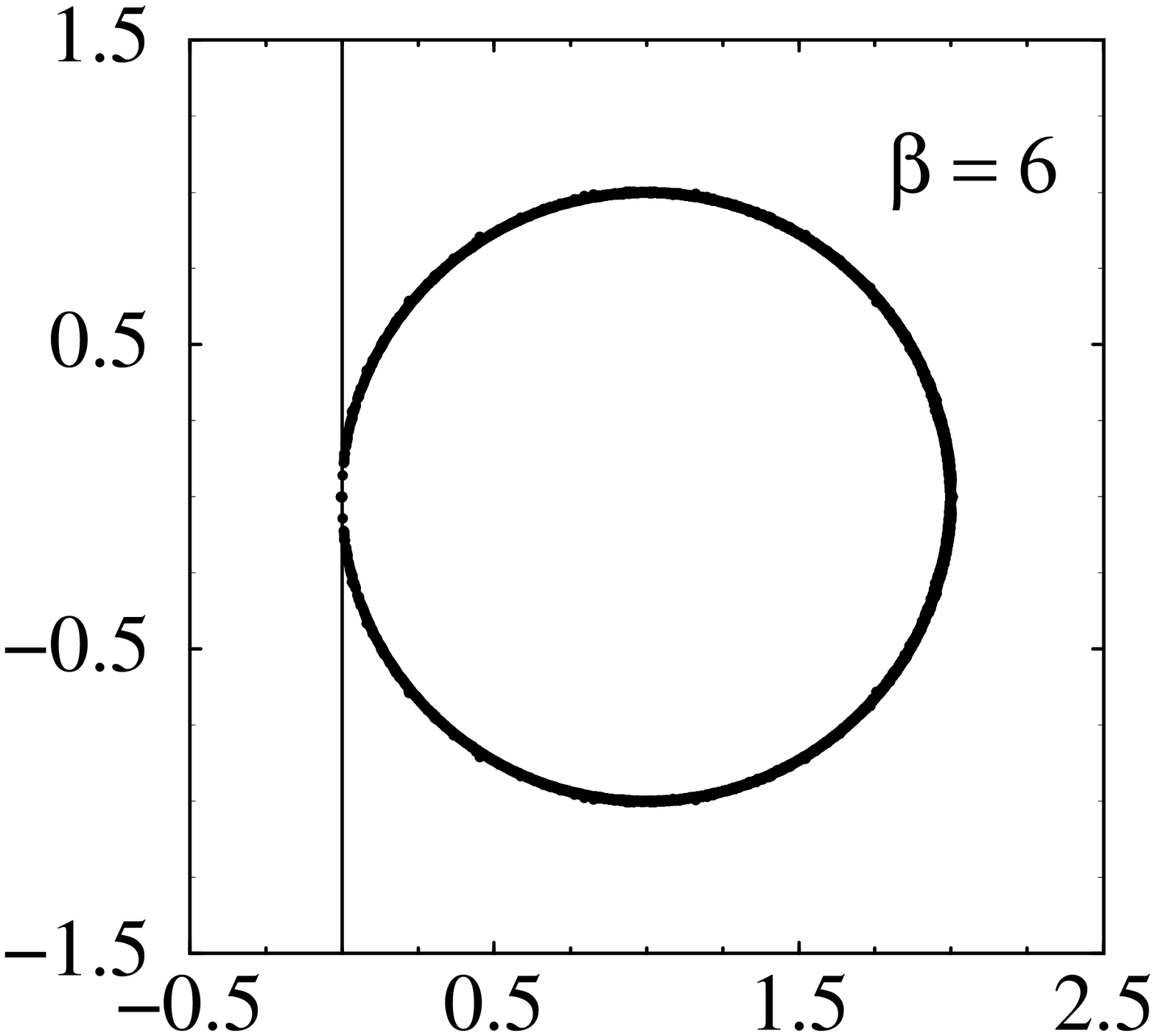}
\end{center}
\caption{Eigenvalues for sets of 25 configurations (superimposed) for 
lattice size $16\times 16$, sampled according to the compact gauge 
action (from Ref.26). \label{fig:FPspectra} }
\end{figure}

For the eigenvalue spectrum of the FP operator we found the situation shown in
Fig.\ref{fig:FPspectra}: The eigenvalues are distributed close to a unit
circle, in particular towards the continuum limit  (growing $\beta$). Towards
the ``hot'' region the fuzziness increases. This demonstrates the predicted
behavior for FP Dirac operators discussed in Sect. \ref{sec:GenProperties}. The
deviation from exact circularity is explained by the approximateness of the
action, which gives raise to fluctuations at smaller $\beta$.

Most remarkable is, that indeed we do find (within the numerical accuracy)
vanishing eigenvalues. These are individual zero modes and computing the matrix
element $\langle v_i|\sigma_3\,v_i\rangle$ from their corresponding
eigenvectors we find values $\pm 1$. Thus these zero modes have definite
chirality. The partner modes with opposite  chirality are at the other edge of
the spectrum, at $\lambda=2$,  irrelevant in the continuum limit. The zero
modes occur whenever the geometric topological charge of the gauge
configuration is non-zero and its value agrees with the number of zero modes,
even at finite $\beta$. This agreement with the ASIT is quantitatively much
better than for the Wilson action.

%%%%%%%%%%%%%%%%%%%%%%%%%%%%%%%%%%%%%%%%%%%%%%%%%%%%%%%%%%%%%%
\subsection{Neuberger's Overlap Dirac Operator}
%%%%%%%%%%%%%%%%%%%%%%%%%%%%%%%%%%%%%%%%%%%%%%%%%%%%%%%%%%%%%%

Motivated by the overlap action\cite{NaNe} 
Neuberger proposed\cite{Ne98} to start with some Dirac operator with sufficiently negative mass,
e.g. the Wilson operator at a value of $m$ corresponding to
$\frac{1}{2D}<\kappa<\frac{1}{2D-2}$
and then construct 
\be\label{DOV}
\DOV = \unitmatrix  + \gamma_5\, \epsilon(\gamma_5\,\DWI) \;.
\ee
Although the actual value of $\kappa$ used in this definition
is largely arbitrary its choice may influence the
approach to scaling in the continuum limit.
The generalized sign function of the hermitian operator
$\gamma_5\,\DWI$  may be interpreted as a realization of 
\be
\frac{\DWI}{|\DWI|}=
\frac{\DWI}{\sqrt{\DWI^\dagger\DWI}}=
\gamma_5 \frac{\gamma_5 \DWI}{\sqrt{(\gamma_5 \DWI)^2}}
\ee
and is determined through the eigenvectors and 
the eigenvalues,
\be\label{defeps}
\epsilon(\gamma_5\,\DWI)= 
U \, \textrm{Sign}(\Lambda)\, U^\dagger \quad \textrm{with}\; 
\gamma_5\,\DWI = U \, \Lambda \,U^\dagger\;.
\ee
($\textrm{Sign}(\Lambda)$ denotes the diagonal matrix of 
signs of the eigenvalue matrix $\Lambda$.)

Gauge configurations with non-zero topological charge imply\cite{Ne98}  exact
zero eigenvalues of $\DOV$; therefore exact chiral modes are realizable. This
was confirmed e.g. in Ref.27; we found an exactly circular eigenvalue spectrum
and zero modes. Due to the freedom of choice of the hopping parameter in the
Wilson action used to construct $\DOV$ it cannot be excluded  that some real
doubler modes of the Wilson action are implicitly mistaken for zero modes
leading to a violation of the ASIT at finite $\beta$. This situation can be
efficiently improved, if one starts already with a better action, e.g. an
approximate  FP action.

An immediate question concerns locality of this operator. In 4D QCD it has been
demonstrated\cite{HeJaLu98} that for large enough $\beta$ one may expect
locality and for a Monte Carlo generated ensemble of gauge configurations
exponential falling off of the effective coupling parameters has been shown.

\begin{figure}[t]
\begin{center}
\epsfxsize=5cm 
\epsfbox{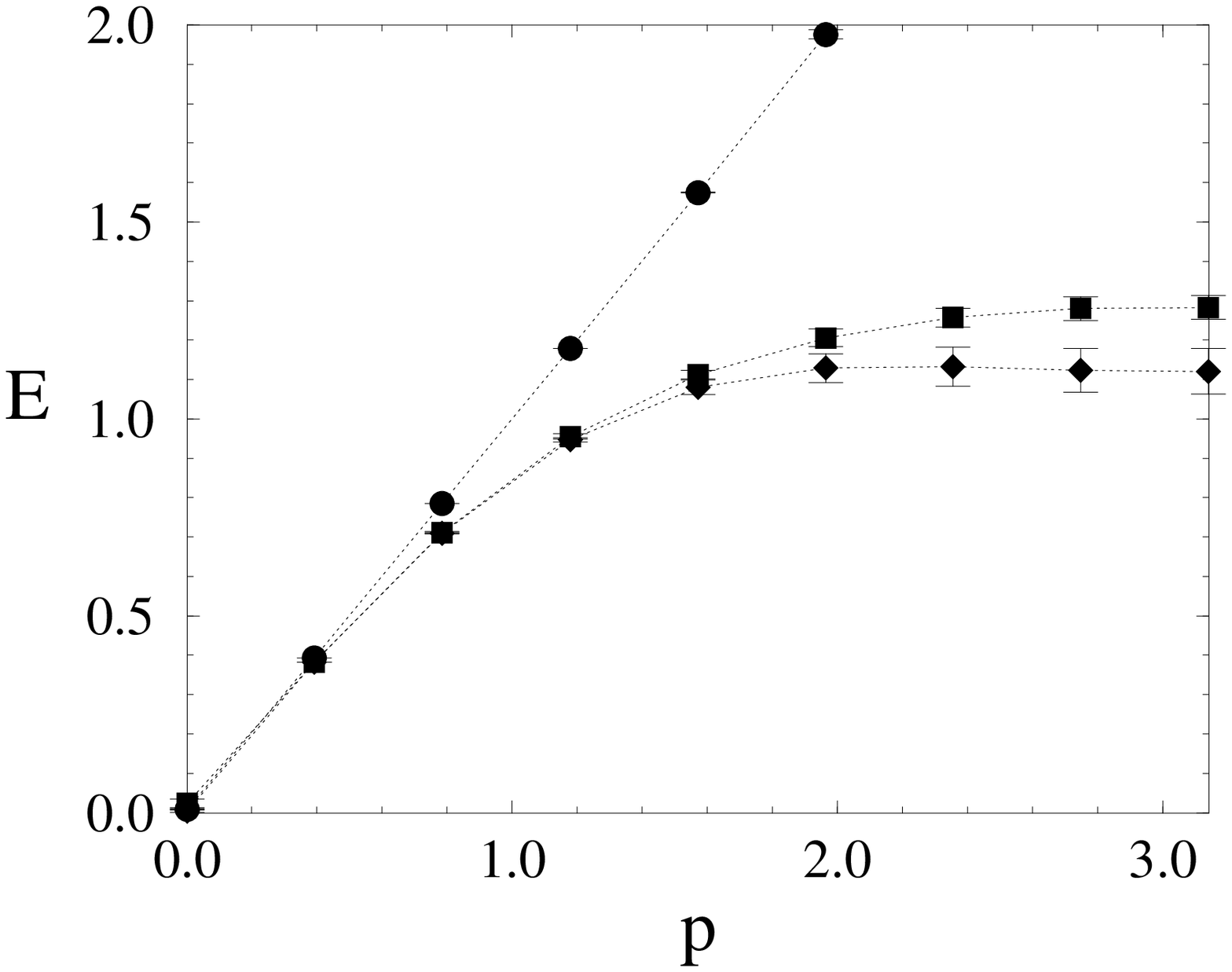}
\epsfxsize=5cm 
\epsfbox{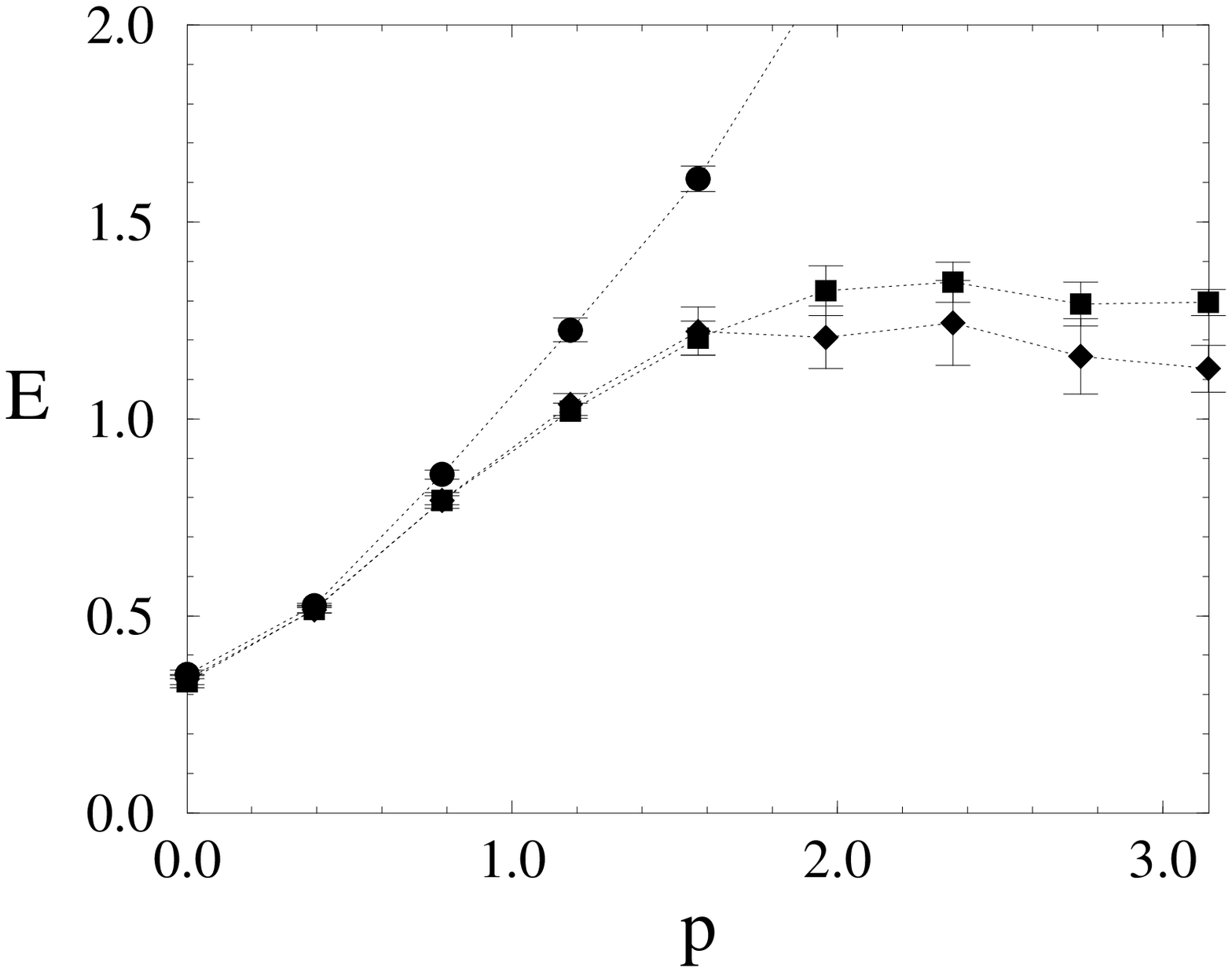}
\end{center}
\caption{The dispersion relations $E(p)$ for the
$\pi$ (left) and the $\eta$ (right) propagators, determined for three actions
discussed in the text: Wilson (squares), fixed point (circles), 
Neuberger (diamonds) (from Ref.27).}\label{fig:PiEtaProp}
\end{figure}

Neuberger's operator provides an explicit example of a GW action with
optimal chirality properties. However, nothing can be concluded as to the 
scaling properties towards the continuum limit. In fact, in Fig.
\ref{fig:PiEtaProp} we see, that the spectral properties of the bound states
propagators are not improved over those of the original Wilson action.  \DOV
is automatically ${\cal O}(a)$ corrected\cite{KiNaNe97,Ni99} and thus
improves scaling for the on-shell quantities; without at least introducing
improvement of the current operators one would not expect improvement for the
propagators, as shown by our results. The two features -- chirality  and
scaling -- seemingly may be considered quite separate issues.  Optimal actions
should allow chiral fermions {\em and} have good scaling properties.

%%%%%%%%%%%%%%%%%%%%%%%%%%%%%%%%%%%%%%%%%%%%%%%%%%%%%%%%%%%%%%
\subsection{Spectral Distribution and chRMT}
%%%%%%%%%%%%%%%%%%%%%%%%%%%%%%%%%%%%%%%%%%%%%%%%%%%%%%%%%%%%%%

The limiting value of the spectral density for small eigenvalues and large
volume,
\begin{equation}
-\pi\,\lim_{\lambda\to 0}\lim_{V\to\infty} \rho(\lambda)=
\langle\bar\psi \psi\rangle\;
\end{equation}
provides an estimate for the chiral condensate due to the Banks-Casher 
relation\cite{BaCa80}.
Exact zero modes are disregarded, only the density close to zero is of
relevance.

In order to study this property of the lattice spectra 
(for the GW Dirac operators)
one should map the eigenvalues from the circular shape to the imaginary axis.
This is done by a stereographic projection
\be
\tilde\lambda = \frac{\lambda}{1-\frac{\lambda}{2}}\;.
\ee
Near zero the resulting distribution on the (tangential) imaginary axis 
agrees with that on the circle. 
The generalization of such a projection for a general GW operator is the
introduction of 
\be
\tilde\D=\frac{\D}{1-R\,\D}
\ee
which anti-commutes with $\gamma_5$, is anti-hermitian and has purely imaginary
spectrum. A procedure to define a subtracted fermion condensate has been
suggested  by Hasenfratz\cite{Ha98c} (within the overlap formalism cf.
Ref.32),
\be
\langle\bar{\psi}\psi\rangle_{\rm sub} = 
-\frac{1}{V}\,\left\langle\textrm{tr}\left(\frac{1}{\D}-R\right)
\right\rangle_\textrm{\footnotesize gauge}=
-\frac{1}{V}\,\left\langle\textrm{tr}\left(\frac{1}{\tilde\D}\right)
\right\rangle_\textrm{\footnotesize gauge}\;,
\ee
(see also Ref.s 12,26,32,33). The expectation value
includes the weight due to the fermionic determinant.

\begin{figure}[tp]
\begin{center}
\epsfig{file=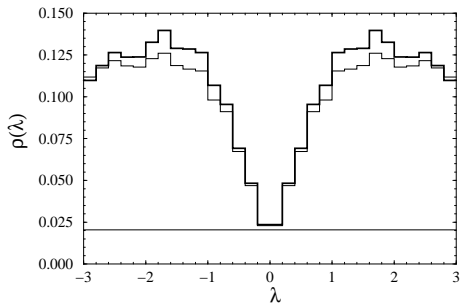,angle=-90,width=6cm}
\end{center}
\caption{\label{fig:CFhisto} The unquenched ($N_f=1$) 
eigenvalue density distribution ($16\times 16$, $\beta=6$)
projected from the circle onto the imaginary axis for 
$\DFP$ (thick lines) and $\DOV$ (thin lines). The horizontal 
line denotes the continuum value at infinite volume.}
\end{figure}

In Fig.\ref{fig:CFhisto} (from Ref.27) the spectral density  for the
Neuberger operator is compared with that of the FP operator; near zero we find good
agreement with each other and with the value expected from the
continuum theory. 

Studying the spectra of the Dirac operators suggests comparison with Random
Matrix Theory\cite{GuMuWe98} (RMT). The spectrum is separated in a fluctuation
part and a smooth background. The fluctuation part is conjectured to follow
predictions in one of three universality classes. For chiral Dirac operators
(chiral  RMT\cite{chRMT}) these are denoted by chUE, chOE and chSE (chiral
unitary, orthogonal or symplectic ensemble,  respectively). Several observables
have been studied in this theoretical context. Comparison of the data should
verify the conjecture and allows one to determine the chiral condensate.  This
information is contained in both, the smooth average of the spectral
distribution, and in the fluctuating part. In particular the distribution for
the smallest eigenvalue $\rho_\textrm{\footnotesize min}(\lambda)$ contains
this observable: Its scaling properties with $V$ are  given by unique functions
of a scaling variable $z\equiv \lambda V \Sigma$, depending on the
corresponding universality class. Usually this is the most reliable approach to
determine $\Sigma$, which then serves as an estimate for the infinite volume
value of the condensate in the chiral limit.

\begin{figure}[tp]
\begin{center}
\epsfig{file=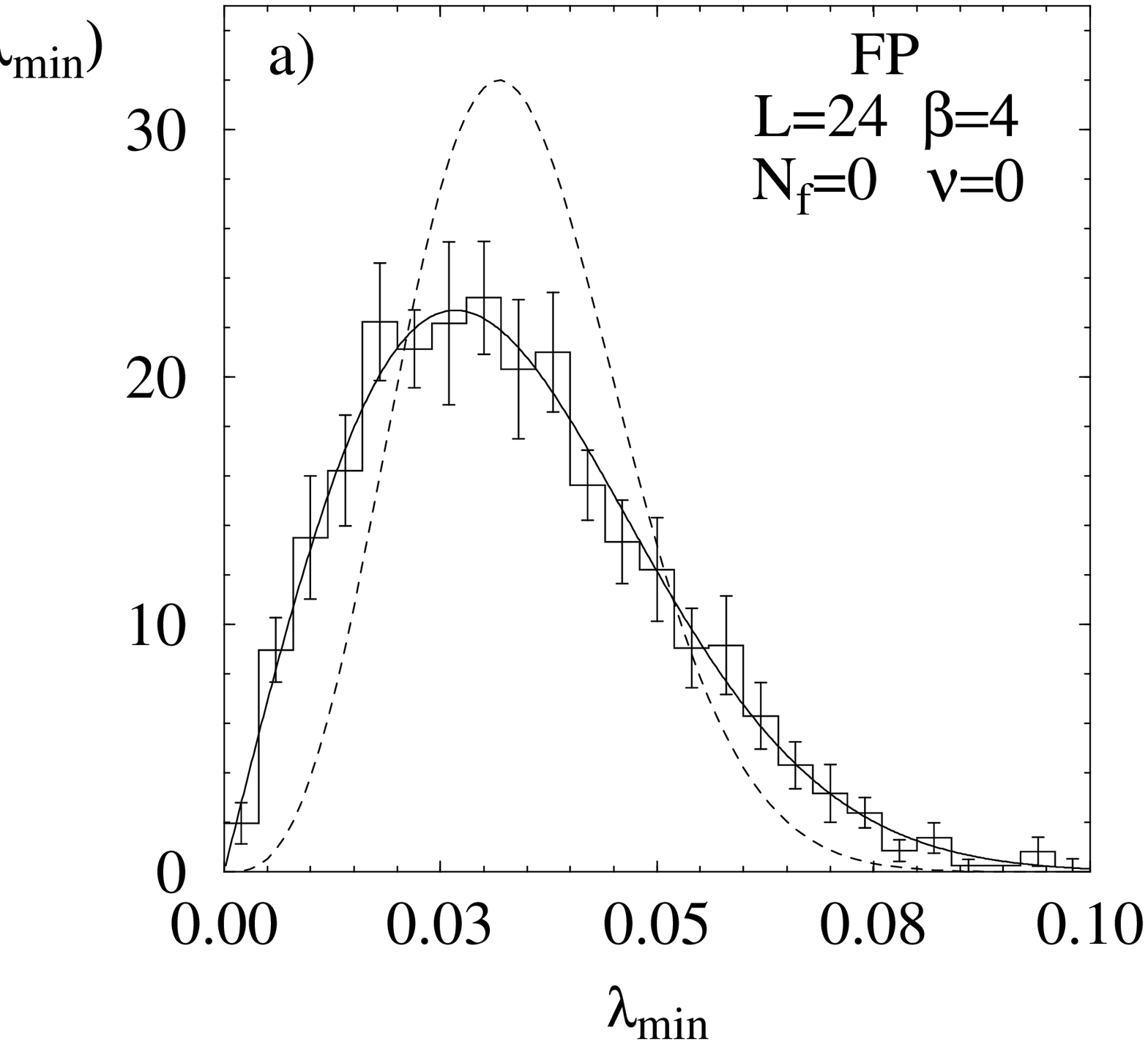,angle=-90,width=4.5cm}
\epsfig{file=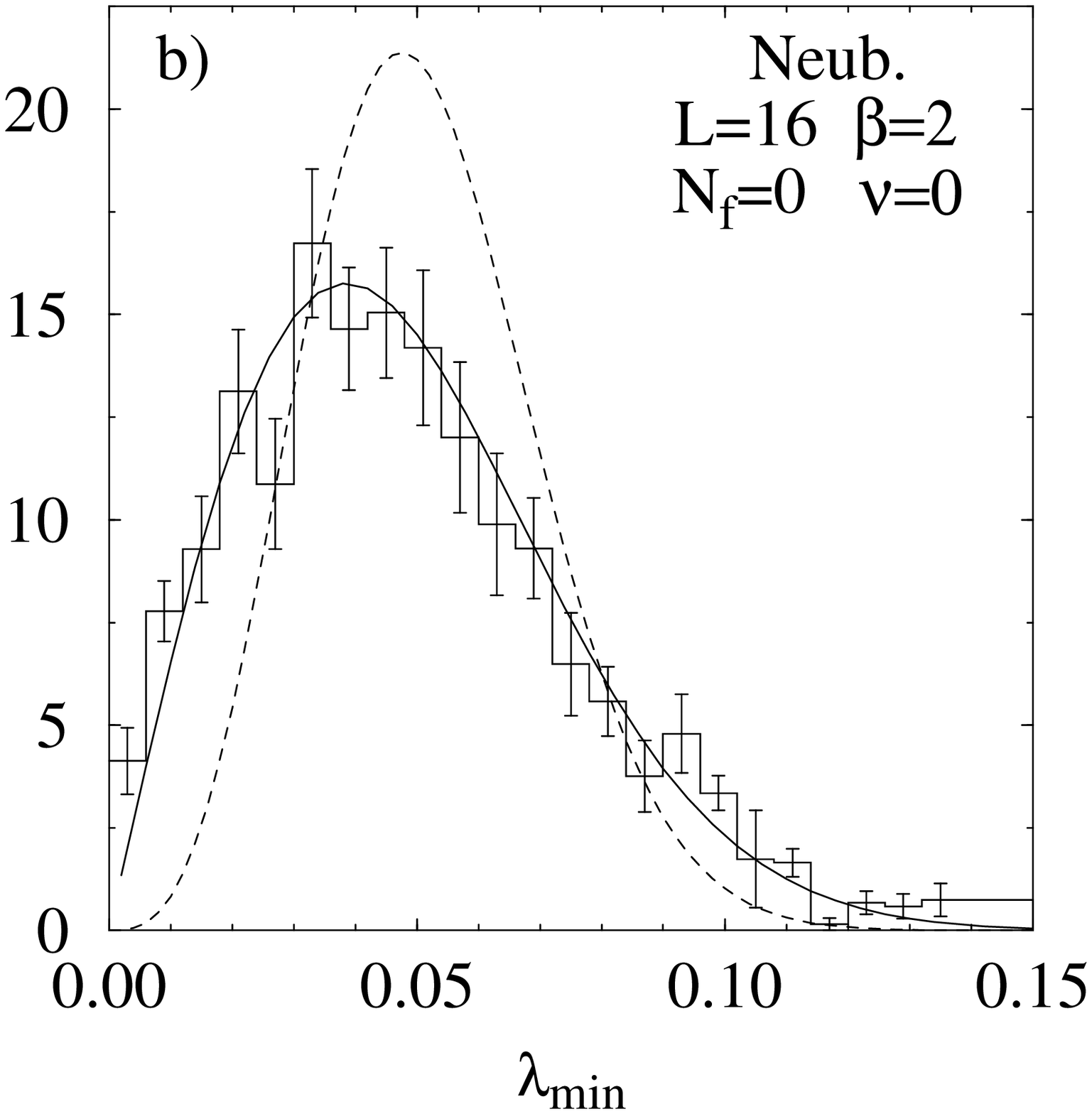,angle=-90,width=4.5cm}
\end{center} 
\caption{\label{fig:chUEPmin} The distribution density for the
smallest eigenvalue in the quenched ($N_f=0$) case in the sector with
topological charge zero. The full line gives the prediction for the chUE class,
the broken line would correspond to the chSE universality class.}
\end{figure}

Within the Schwinger model we found\cite{FaHiLa99} the universal properties
of the (expected) chUE-class, unless the physical lattice volume is too small.
In Fig. \ref{fig:chUEPmin} we show the distribution of the smallest
eigenvalue for two of the actions studied.  Further examples, also including
dynamical fermions and 4D applications, are discussed in
Ref.s\cite{FaHiLa99,Be,EdHeKi99}. 
It has become evident, that chRMT describes the fluctuating part of
the spectrum satisfactorily within the expected universality classes.  It
provides a means to separate the universal features from quantities like  the
chiral condensate, which have dynamic origin. Or, to say it 
(tongue in cheek) more provocative, to
separate the known thus uninteresting, universal properties from the unknown,
physical properties of the system.

%%%%%%%%%%%%%%%%%%%%%%%%%%%%%%%%%%%%%%%%%%%%%%%%%%%%%%%%%%%%%%
\section{Conclusion}
%%%%%%%%%%%%%%%%%%%%%%%%%%%%%%%%%%%%%%%%%%%%%%%%%%%%%%%%%%%%%%

The main messages I wanted to deliver here are
\begin{itemize}
\item
Lattice gauge theory has made a big step forward in the understanding of 
chiral fermions and their lattice formulation. The Ginsparg-Wilson  relation
is a cornerstone in that development. Meanwhile U(1)  chiral gauge theories
with anomaly-free multiplets of Weyl fermions have been constructed 
properly\cite{Lu99}.
\item
Neuberger's overlap action is a wonderful testing ground for chirality;
its implementation in 4D is computationally expensive.
Perfect actions ideally provide the best of both, chirality and scaling, but
their construction in realistic 4D models seems to have unsurmountable  problems.
In a 2D gauge theory, however, we were able to construct an explicit example,
which  has all the beautiful features expected. 
\item
The spectra of Dirac operators have universal features, which are described by
chiral RMT. This opens the path to efficient methods to extract from the data
non-perturbative quantities like the chiral condensate. 
\item
The Schwinger model has provided an excellent test case to study various 
features also observed in QCD. Although maybe of different dynamical origin,  
mechanisms like confinement, chirality and symmetry breaking can be nicely
studied in their lattice realization. This may lead to a better understanding
also for the 4D theories.
\end{itemize}

%%%%%%%%%%%%%%%%%%%%%%%%%%%%%%%%%%%%%%%%%%%%%%%%%%%%%%%%%%%%%%
\section*{Acknowledgments}
%%%%%%%%%%%%%%%%%%%%%%%%%%%%%%%%%%%%%%%%%%%%%%%%%%%%%%%%%%%%%%
Ludwig Streit has been a colleague and friend since long; I learned a lot from
him although we never  collaborated (except for surfing in Santa Barbara). I
want to thank F. Farchioni, C. Gattringer, I. Hip, T. Pany and M. Wohlgenannt
for collaboration on various topics within the Schwinger model and for allowing
me to present also some of their contributions in this text. This write-up
tries to present the results available at the conference in Lisbon, Oct. 1998;
only the references have been brought up to date. Support by Fonds 
zur F\"orderung der Wissen\-schaft\-lichen Forschung 
in \"Osterreich, Project P11502-PHY, is gratefully acknowledged.

%\bibliographystyle{prsty} %unsrt npb
%\bibliography{/u1/people/cbl/refs/lgt+1990,/u1/people/cbl/refs/lgt-1989}
%5 \cite{Ha98c}
%26 \cite{FaLaWo98}
%27 \cite{FaHiLa98}
%32 \cite{Ne98d}
%12,26,33,32\cite{Ne98,FaLaWo98,ChZe99,FaHiLa99}

\end{document}